\documentclass[12pt]{iopart}

\usepackage{graphicx}
\begin{document}

\title{Entropy of the holographic dark energy and generalized second law}

\author{Praseetha P.  and Titus K. Mathew}

\address{Department of Physics, Cochin University of Science and Technology,Cochin, India}
\ead{praseethapankunni@cusat.ac.in,titus@cusat.ac.in}
\begin{abstract}
In this paper we have considered the holographic dark energy and studied it's cosmology and thermodynamics. We have analysed the generalized 
second law (GSL) of thermodynamics in a flat universe consists of interacting dark energy and dark matter. We did the analysis both under thermal equilibrium and non-equilibrium 
conditions. If the apparent horizon is taken as the boundary of the universe, we have shown that the rate of change of the total entropy of the universe is 
proportional to $(1+q)^2,$ which in fact shows that the GSL is valid at the apparent horizon irrespective of the sign of $q,$ the deceleration parameter. Hence for any form 
of dark energy the apparent horizon can be considered as a perfect thermodynamic boundary of the universe. We made a confirmation of this conclusion by using the holographic dark energy. When event horizon 
is taken as the boundary, we found that the GSL is only partially satisfied. The analysis under non-equilibrium conditions revealed that the GSL is satisfied if the dark energy 
temperature is greater than the temperature of dark matter.
\end{abstract}

\maketitle

\section{Introduction}
The observational evidence have proved beyond doubt that the present universe 
is accelerating \cite{Perl1,Shariff1,Koivisto1,Daniel1,Fedeli1}.
The exotic matter, the dark energy, causing
the accelerated expansion of the universe is assumed to have a
negative pressure.
Various models were proposed aimed at understanding the
properties of dark energy. Among these, the cosmological constant 
is the prominent one, but it faces the problems of fine tuning 
and also unable to explain the coincidence of dark energy density and dark matter density.
This leads to consideration of dynamic dark energy models. 
Among which 
holographic dark energy model
\cite{Xang1,Huang1,Xang2,Xang3,Setare1} based on the holographic principle 
is one of the prominent models discussing in the recent literature.
According to the holographic principle
\cite{Skind1}, the vacuum energy density is bounded.The
significance of the principle lies in the constraint that the  total
energy inside a region of size L, should not exceed the mass of a
black hole of the same size. From effective quantum field theory, an
effective IR cut off can saturate the length scale, so that the dark
energy density (the vacuum energy density) can be written as
\begin{equation}
\rho_{\Lambda}=3c^{2}(M_{p}^{2}) L^{-2},
\end{equation} where c is a
dimensionless numerical factor.The possible choices for the
IR cut-off are the Hubble horizon distance, particle horizon distance, event horizon distance
and some generalized IR cut-off \cite{Miaoli1,Malekjani1,Huang2}. But the first two options could not
support the accelerated expansion of the universe while the event
horizon posed the problem of causality.Thus a new model has been 
introduced by Granda and
Oliveros \cite{Granda1}by taking the Ricci scalar as the IR cut-off,  and later studied by many others \cite{tkm2}. This is seemed to be 
efficient in solving the coincidence problem,
causality problem, and also the fine tuning problem. 
A modified form of this model was studied in interaction with
dark matter assuming a hidden non-gravitational coupling existing
between them \cite{Chimento1,Chimento2,Ujjal1}.
In this paper we will explore the thermodynamics, especially the status of the generalized second law 
of thermodynamics in a flat universe with holographic dark energy with 
Ricci scalar as the IR cut-off, interacting with the dark matter.

Thermodynamics of cosmological models is of great interest after the
discovery of the black hole thermodynamics \cite{Ujjal1} by
Bekenstein and Hawking \cite{Davies1} which says that the entropy of
the black hole is proportional to the area of it's event horizon  \cite{Wang1,Gong1}. Through the Hawking radiation,
the black hole could lose it's thermal energy which results in the decrease
of its entropy and this is not allowed by the second
law if black hole is assumed as an isolated system. 
It is explained as even though
the entropy of the black hole decreases the loss is compensated by
the increase in the entropy of the matter outside the black hole
boundary, eventually the total entropy of the black hole plus that of the 
surrounding is increasing which is termed as the generalized second law
(GSL) \cite{Bek1,Bek2}.The black hole thermodynamics is applied
in the cosmological holographic model by L.Susskind \cite{Skind1}and 't Hooft \cite{G'tHooft1}
which says that the maximum entropy of a bounded region is
proportional to its area rather than its volume. Gibbons
and Hawking \cite{Gibbons1} studied the de-Sitter model of the universe where apparent horizon and the
cosmological event horizon are coincide with each other and
found that the GSL is satisfied. 
Assuming the whole universe as
a thermodynamic system in which the validity of the first law of
thermodynamics is undoubtedly assured, the present paper quests for
the validity of the generalized second law of
thermodynamics(GSL), in a universe filled with holographic dark energy interacting 
with the dark matter present. The first
law of thermodynamics and GSL is found to be satisfied for the
apparent horizon while violated in the case of event
horizon for dark energy with constant equation of state and in a universe with generalized Chaplygin gas \cite{Gong1}. 
The event horizon describes a global concept of
space-time, hence a difficulty is said to appear in understanding
Hawking radiation clearly \cite{Cai1}.Thus there are arguments in
favor of apparent horizon as the real physical boundary
\cite{Gong1}. Hence we consider whether the GSL is
satisfied by the apparent horizon and event horizon in the realm of 
holographic dark energy model, in order to discern the real
physical boundary of the universe from the point of view of thermodynamics.

\section{Interacting dark energy model}

The Friedmann equations for flat universe with FRW metric can be
written as
\begin{equation}
\label{eqn:friedmann1} 3H^{2}=\rho_{m}+\rho_{de}
\end{equation}
where H is the Hubble parameter, $\rho_{m}$ is the energy density of
the dark matter and $\rho_{de}$ is that of the holographic
dark energy. The expression for the holographic 
dark energy with Ricci scalar as the IR cut-off is
\begin{equation}
\label{eqn:mhrde} \rho_{de}=2(\dot{H}+(3/2)\alpha H^{2})/
\Delta
\end{equation}
where $\dot{H}$ represent the derivative of H with respect to the
cosmic time t and $\triangle = \alpha- \beta$
 where $\alpha$ and $\beta$  are parameters of the model.
The interaction between the dark energy and dark matter is studied
through the conservation equations given below
\begin{equation}
\dot{\rho}_{de}+3H(\rho_{de}+p_{de})=-Q
\end{equation}
\begin{equation}
\dot{\rho}_{m}+3H\rho_{m}=Q
\end{equation}
where $p_{de}$ is the pressure of the dark energy, $Q$ is the
interaction term, dot represents the derivative with  respect to time. Dark
matter is assumed to be pressure less. The term $Q$ can have mainly
three forms, $Q=3b H (\rho_{de}+\rho_m)$, $Q=3b H \rho_{m}$ and
$Q=3b H \rho_{de}$ \cite{TFFu,SChat}. For the present study we have
considered the form, $Q=3b H (\rho_{m}+\rho_{de})$.

\begin{figure}[ht]
\centering
\includegraphics[scale=1]{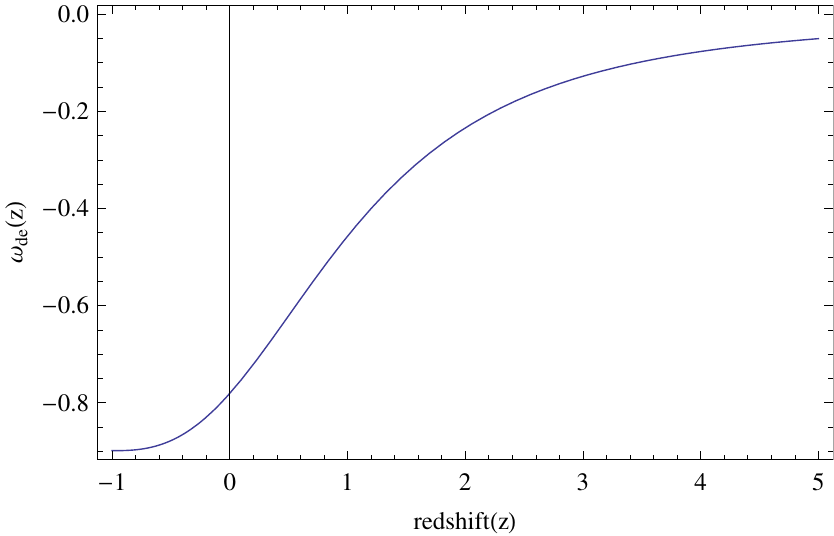}
\caption{The evolution of the equation of state parameter of the
interacting holographic dark energy for the parameter $(\alpha,\beta)=(1.2,0.1)$ }
\label{fig:eos}
\end{figure}

Substituting, the dark energy density given by equation
(\ref{eqn:mhrde}) in the Friedmann
equation(\ref{eqn:friedmann1}), and changing the variable cosmic time
t to the variable $x=\log a$ and differentiating the
equation once more leads to 
the second order
 differential equation,
\begin{equation}
\label{eqn:diff} {d^2h^2 \over dx^2}+3\left(\beta +1 \right) {dh^2
\over dx}+ 9\left(\beta + b \Delta \right)h^2=0
\end{equation}
where $h=H/H_0$, $H_0$ is the present value of Hubble parameter. The
general solution for this can be obtained as
\begin{equation}
h^2=c_1e^{\frac{3}{2} m_1 x}+c_2e^{\frac{3}{2} m_2x}
\end{equation}
where the constants are evaluated using the initial
conditions, $h^2|_{x=0}=1$ and ${dh^2\over
dx}|_{x=0}=3\Omega_{de0}\Delta-3\alpha$, are obtained as
\begin{equation}
c_1={2\left(\Omega_{de0} \Delta -\alpha \right) - m_2 \over m_1-m_2}
, \,\,\,\,\,\,\,\,\,\, c_2=1-c_1
\end{equation}
and $m_{1,2}$ are obtained as
\begin{equation}
 m_{1,2} = -1 -\beta \mp \sqrt{1-4b \alpha - 2\beta + 4 b \beta + \beta^2}.
\end{equation}
Using the Friedmann equation, the dark energy density parameter can
be obtained as 
\begin{equation} 
\Omega_{de}=c_1e^{\frac{3}{2} m_1 x} 
+c_2e^{\frac{3}{2}m_2 x}-\Omega_{m0}e^{-3x}.
\end{equation}
The pressure of the Dark Energy  and its equation of state parameter
are calculated as,
\begin{equation}
P_{de}=-\left[c_1\left(1+\frac{m_1}{2}  \right) e^{\frac{3}{2}m_1 x} + 
c_2\left(1+\frac{m_2}{2}  \right) e^{\frac{3}{2}m_2 x} \right].
\end{equation}
and
\begin{equation}
\omega_{de} = -1- \frac{1}{2} \left({c_1 m_1 e^{\frac{3}{2}m_1 x} + c_2 m_2 e^{\frac{3}{2}m_2 x} + 2 \Omega_{m0} e^{-3x} 
\over c_1  e^{\frac{3}{2}m_1 x} + c_2 e^{\frac{3}{2}m_2 x} -  \Omega_{m0} e^{-3x} } \right).
\end{equation}
The evolution of the equation of state of the interacting holographic dark energy is plotted
against redshift $z$ for the model parameters,
$(\alpha,\beta)=(1.2,0.1)$ and is shown 
in figure(\ref{fig:eos}). The figure shows that in the past stage of the universe the equation of state was nearly zero. As 
the universe evolves the equation of state decreases and stabilizes as $z \rightarrow -1.$
In a dark energy dominated universe, $\Omega_{de0} \sim 1$ and $\Omega_m \sim 0$, the coefficients in the above equation become $m_1=-2, \, m_2=-2\beta, \, c_1=0$ and $c_2=1.$
Then the equation of state parameter become $\omega_{de}=-1+\beta$, which shows that for positive values of $\beta$ the equation of state $\omega_{de}>-1$ 
corresponds to quintessence type behavior and for $\beta<0$ the equation of state $\omega_{de}<-1$ corresponds to phantom behavior.

The deceleration parameter $q$ for the dark energy considered is obtained as 
\begin{equation} \label{eqn:qpar}
q=-{3 \left(c_1 m_1 e^{\frac{3}{2} m_1 x} + c_2 m_2 e^{\frac{3}{2} m_2 x} \right) \over 
4 \left(c_1 e^{\frac{3}{2} m_1 x} + c_2  e^{\frac{3}{2} m_2 x} \right) } - 1.
\end{equation}
The nature of deceleration parameter is studied by plotting $q$
\begin{figure}[ht]
\centering
\includegraphics{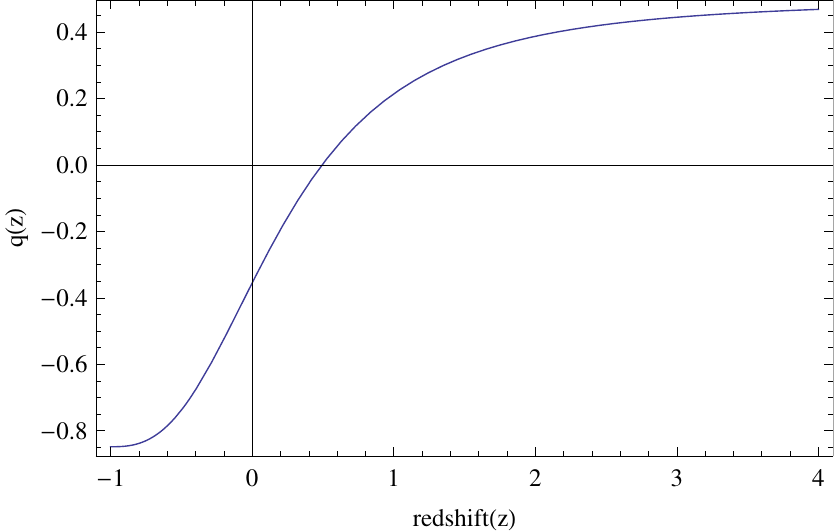}
\caption{The study of deceleration parameter of the interacting dark energy for the model
parameter  $(\alpha,\beta)=(1.2,0.1)$ } \label{fig:q}
\end{figure}
against the redshift $z$ and the result is shown in the
figure \ref{fig:q}. It is evident from the figure that during the initial times in the evolution of the universe the $q$ 
parameter was positive implies deceleration in the expansion. As the universe expands the $q$ parameter enters the negative value region
in the recent past, implying that the current acceleration was started in the recent past. It is also seen that the $q$ parameter
saturates as $z \rightarrow -1.$
For dark energy dominated universe with $\Omega_m \sim 0$, the deceleration parameter become $q=-1+(3/2)\beta.$ This implies that $\beta>0$ corresponds to 
quintessence type behavior, $q >-1$ and $\beta<0$ corresponds to phantom behavior, $q<-1.$

\section{GSL under thermal equilibrium condition}
In this section the validity of the generalized second law of thermodynamics of
the dark energy under thermal equilibrium is studied.  The validity of the GSL
implies that the sum of the change in entropy of the dark sectors added
with the entropy change of the cosmological horizon is greater than or equal to zero
\cite{Setare1}. That is,
\begin{equation}
\dot{S}+\dot{S_{h}}\geq 0
\end{equation}
where $\dot{S}$ denotes the change in entropy of dark sectors with
respect to cosmic time and $\dot{S_{h}}$ is the change in horizon
entropy with respect to cosmic time.
 The study is done by considering 
that the dark sectors and the horizon were in  thermal
equilibrium so that temperature of the dark sectors is equal to the temperature of the horizon.
 In the following we will analyse separately the GSL for universe with 
apparent horizon as boundary and event horizon as the boundary. 

\subsection{GSL validity inside apparent horizon}

Here the validity of the GSL is that sum of entropy of the dark energy, dark matter and the
apparent horizon must increase with time \cite{Karami1,Karami2}. Apparent horizon do exist for all kinds of FRW universes \cite{Cai2}.
The apparent horizon is a causal horizon for the dynamical space time and is associated with gravitational entropy and 
surface gravity \cite{Hayward1,Hayward2,Bak1}. For a universe with FRW metric the apparent horizon radius can be obtained as
\begin{equation}
 r_{a} = {1 \over \sqrt{H^2 + k/a^2}}.
\end{equation}
For flat FRW universe ($k=0$), the apparent horizon become $r_a = 1/H,$ which is same as the radius of the Hubble horizon.
The temperature on the apparent horizon can be defined as 
$ T_a = |\kappa|/2\pi$, where $\kappa$ is the surface gravity \cite{Cai2}.
For flat universe the surface gravity depends on the horizon radius, and simple considerations will leads to the 
temperature as
\begin{equation}
T_{a}=\frac{H}{2\pi}.
\end{equation}

The entropy of the apparent horizon is  $S= \frac{A}{4G}$ with $8\pi G=1$
\cite{Gibbons1,Padm1,Davies2,Izquierdo1,tkm1}, where the area $A=4\pi r_a^2.$ Thus the entropy of the apparent horizon can be obtained
as \cite{Davies1,Davies2}
\begin{equation}
S_{a}=\frac{8\pi^2}{H^{2}}
\end{equation}
The entropy of the dark sectors can be found using Gibbs' equation
given below
\begin{equation}
TdS=dE+PdV
\end{equation}
where $V=\frac{4}{3}\pi r_{a}^{3}$ is the volume occupied by the dark
entities and $E=\frac{4}{3}\pi r_{a}^{3}(\rho_{de}+\rho_{m})$ and
the pressure $P$ has contribution only from the dark energy since
dark matter is non-relativistic hence pressureless. Thus using the above equations and the
conservation equation of the dark sectors given in the previous
section, the total entropy variation with respect to $x=\log a$ is obtained as
\begin{equation} \label{eqn:S1}
S^{'}=\frac{16 \pi^2}{H^{2}}+\frac{16 \pi^2}{H^{2}}\left(1+\frac{3}{2} 
(1+\omega_{de} \Omega_{de})\right)q
\end{equation}
where $\omega_{de}$ is the equation of state parameter of the dark
energy, $\Omega_{de}$ is the dark energy density parameter and $q$
is the deceleration parameter.
The GSL is valid if $S^{'} \geq 0.$ Since $H^2>0$, the validity of the GSL requires that, $q \geq -1/[1+ (3/2) (1+\omega_{de} \Omega_{de})].$
This condition can approximately be translated as  $H \geq 1/t(1-1/c),$ where $c \sim (3/2)(1+ \omega_{de} \Omega_{de})$ (a comparatively 
smaller value). In the present case where $c$ is comparatively small hence $H\geq 1/t.$
This is a reasonable bound for $H$ in this context
and will lead to the increase in the change of entropy hence the validity of the GSL. 

The equation (\ref{eqn:S1}) can be further simplified using the relation for the $q$ parameter, 
\begin{equation}
 q = -1 - \frac{\dot{H}}{H^2} = \frac{1}{2}\left(1+3\omega_{de} \Omega_{de} \right).
\end{equation}
The equation (\ref{eqn:S1}) can be re-written as 
\begin{figure}[ht]
\centering
\includegraphics[scale=1]{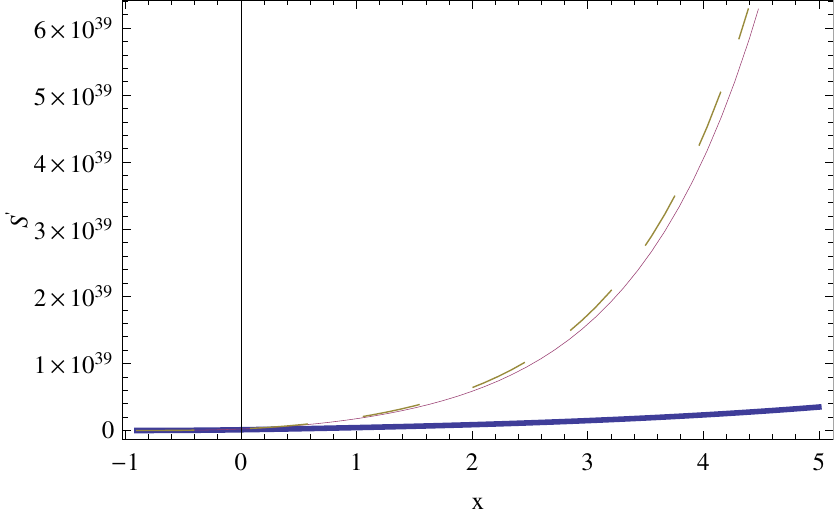}
\caption{The behavior of $S^{'}$ for the parameters
$(\alpha,\beta)=(1.2,0.1)$ thick continuous line,
$(\alpha,\beta)=(1.3,0.3)$ thin continuous
line,$(\alpha,\beta)=(1.4,0.3)$ dashed line and
 with the interaction
coupling constant b=0.001 inside apparent horizon under thermal
equilibrium conditions} 
\label{fig:Svar1}
\end{figure}
\begin{figure}[ht]
\centering
\includegraphics[scale=1]{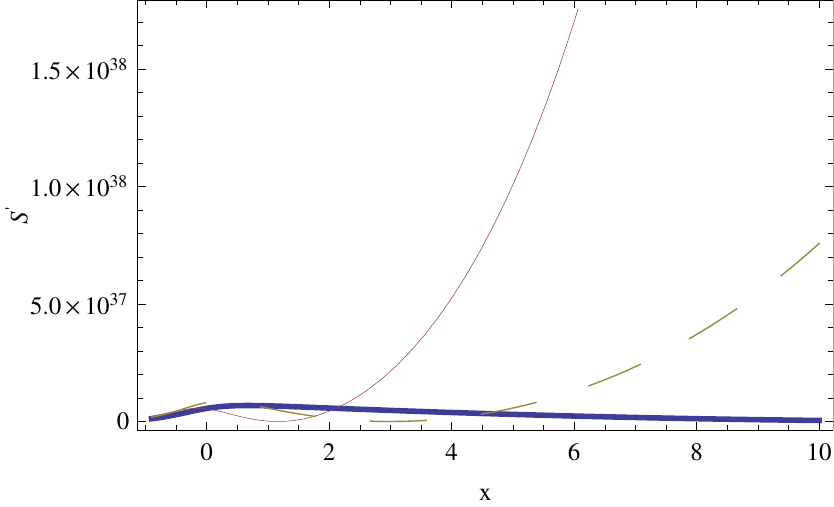}
\caption{The behavior of $S^{'}$ for the parameters
$(\alpha,\beta)=(1.01,-0.01)$ thick continuous line,
$(\alpha,\beta)=(1.2,-0.1)$ thin continuous
line,$(\alpha,\beta)=(1.3,-0.05)$ dashed line and
 with the interaction
coupling constant b=0.001 inside apparent horizon under thermal
equilibrium conditions} 
\label{fig:Svar2}
\end{figure}
\begin{equation} \label{eqn:sprime1}
 S^{'}={16\pi^2 \over H^2} \left(1 + q \right)^2 = {36\pi^2 \over H^2} \left(1+\omega_{de} \Omega_{de} \right)^2
\end{equation}
This equation shows that, as far as  $H>0,$ the GSL is satisfied for almost all kinds of dark energy, irrespective of the sign of $q$ which is positive for decelerated expansion (radiation or matter dominated era) and 
negative for accelerated expansion of the universe. Deceleration parameter $q>-1$ (corresponds to $\omega_{de}>-1$) in the quintessence phase and 
$q<-1$ (corresponds to $\omega_{de}<-1$) in the phantom phase of expansion.  
So both in the quintessence phase of expansion and in the phantom phase the GSL is satisfied with apparent horizon as the boundary of the universe. 
Current cosmological observations hint that $\omega_{de}$ may be as lower as -1.5 \cite{Knop1}. 
 In the dark energy 
model we have considered equation (\ref{eqn:qpar}) implies that, in dark energy dominated case the universe is in the quintessence phase when $\beta>0$ and when 
$\beta <0$ the universe enter 
the phantom phase of expansion as $z \rightarrow -1.$  Equation (\ref{eqn:sprime1}) shows that the 
 GSL is satisfied for both positive and negative values of $\beta$ parameter.

We have studied the evolution of $S^{'}$ with $x$ using the equation of $q$ derived in the last section to check the above conclusions. 
 The plot \ref{fig:Svar1} showing the behavior of the
$S^{'}$ using the equation (\ref{eqn:S1}) for various choices of the model parameters $(\alpha,\beta)$ and it shows that 
the GSL is satisfied always in conformation with equation (\ref{eqn:sprime1}).

\subsection{GSL validity inside event horizon}

In this section we analyse the GSL with event horizon as the boundary under thermal equilibrium condition.
The event horizon distance can be evaluated using the standard relation,
\begin{equation}
R_h=\frac{1}{1+z} \int_z^{-1} \frac{dz}{H}.
\end{equation} 
On integration the event horizon is turn out to be
\begin{equation}
 R_h ={ K_1 (1+z)^{3m_2+1}  \over \sqrt{c_1 (1+z)^{(3/2)(m_2-m_1)}}} \, {_2}F_1[g_{+},0.5,1+g_{+}, \frac{c_1}{c_2}(1+z)^{(3/2)(m_2-m_1)}]
\end{equation}
for positive $\beta$ values and 
\begin{equation}
 R_h={K_2 \over \sqrt{c_2}} (1+z)^{3m_2+1} {_2}F_1[g_{-},0.5,1+g_{-}, \frac{c_1}{c_2}(1+z)^{(3/2)(m_2-m_1)}]
\end{equation}
for negative values of $\beta.$ The functions ${_2}F_1[...]$ are the hypergeometric functions. The constants $K_1, K_2, g_{+}, g_{-}$ are different for different values of $\beta,$ and the suffix ${+}({-})$ refers to positive (negative) values of
$\beta.$ The values of the various constants is found to be $g_1=0.185, 0.238, 0.238$ for $(\alpha,\beta)=(1.2,0.1),(1.2,0.3),(4/3,0.3)$ respectively and $K_1=8.484 \times 10^{17}$ for 
all the parameter sets. For negative values of $\beta$ the constants are found to be $g_2=0.345,0.349, 0.349$ for $(\alpha,\beta)=(1.01,-0.01),(1.2,-0.1),(4/3,-0.1)$ respectively and 
$K_2=-4.227 \times 10^{17}.$

In this case the temperature is $T=1/4\pi R_h$ and area of the horizon is $A=4\pi R_h^2.$ Using the Gibb's relation 
the rate of change of entropy of dark energy plus matter is
\begin{equation}
 T \left(S_{de}^{'}+S_m^{'} \right) = H^{-1} \left(\rho_{de}+\rho_m + p_{de} \right) 4\pi R_h^2 \left(\dot{R_h} - H R_h \right).
\end{equation}
\begin{figure}[ht]
\centering
\includegraphics[scale=1]{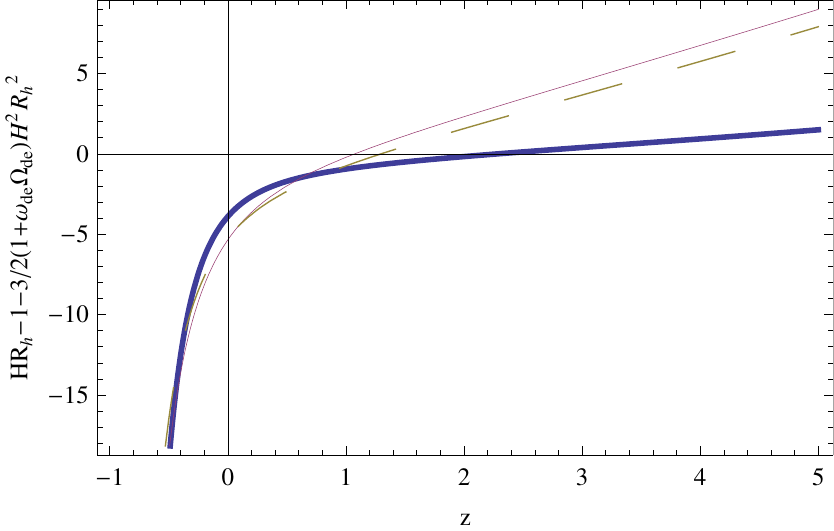}
\caption{The behavior of $S^{'}$ for the parameters
$(\alpha,\beta)=(1.2,0.1)$ thick continuous line,
$(\alpha,\beta)=(1.3,0.3)$ thin continuous
line,$(\alpha,\beta)=(4/3,0.3)$ dashed line  with the interaction
coupling constant b=0.001 inside event horizon under thermal
equilibrium conditions}
\label{fig:ehpob}
\end{figure}
\begin{figure}
\centering
\includegraphics[scale=1]{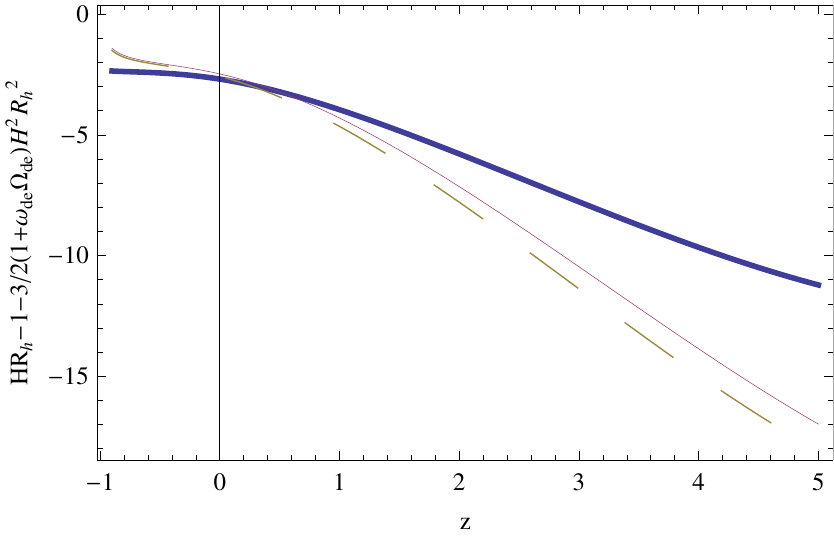}
\caption{The behavior of $S^{'}$ for the parameters
$(\alpha,\beta)=(1.01,-0.01)$ thick continuous line,
$(\alpha,\beta)=(1.2,-0.1)$ thin continuous
line,$(\alpha,\beta)=(4/3,-0.1)$ dashed line, with the interaction
coupling constant $b=0.001$ inside event horizon under thermal
equilibrium conditions}
\label{fig:ehneb}
\end{figure}
Substituting the temperature $T$ and using the relation $\dot{R_h}=HR_h-1,$\cite{Karami2} the above equation become,
\begin{equation}
 \left(S_{de}^{'}+S_m^{'} \right) = - H^{-1} \left(\rho_{de}+\rho_m + p_{de} \right) 8\pi^2 R_h^3.
\end{equation}
Considering that the fluid inside the horizon satisfies the dominant energy condition $(\rho + P) > 0$ \cite{Davies1,Davies2}, the above equation shows that the rate 
of change of entropy of dark energy plus dark matter within the event horizon decreases as far as $R_h>0.$ Adding the rate of change of horizon entropy to the above equation 
we get the rate of change of the total entropy as 
\begin{equation} \label{eqn:S2}
 S^{'}=H^{-1} \left[16\pi^2 R_h \left(\dot{R_h} - \frac{R_h^2}{2}  (\rho_{de}+\rho_m+p_{de}) \right) \right]
\end{equation}
The above equation can be modified by using $\dot{H}=-(1/2)(\rho_{de}+\rho_m+P_{de}),$ we get
\begin{equation}
 S^{'}=H^{-1} \left[ 16\pi^2 R_h \left(\dot{R_h} + \dot{H} R_h^2 \right) \right].
\end{equation}
For $H>0$ and $R_h>0$ the  GSL is satisfied i.e. $S^{'} \geq 0$ if $\dot{R_h}+\dot{H}R_h^2 \geq 0.$ This can integrated to, if one avoid the constant of integration, as
$HR_h \geq 1.$ The exact condition for the validity of GSL become 
\begin{equation}
 \dot{R_h} \geq \frac{1}{2} \left(\rho + p \right) R_h^2
\end{equation}
This shows that the GSL is valid only if the universe satisfies the dominant energy condition $(\rho + p) \geq 0$ \cite{Davies1,Davies2} in the case of increasing horizon radius, 
i.e. in the quintessence phase.

Using the Friedmann equation and 
conservation equation 
 a more exact relation for the validity of GSL from the condition $\dot{R_h}+\dot{H}R_h^2 \geq 0,$ can be obtained as 
\begin{equation} \label{eqn:horcond1}
 HR_h \geq 1+\frac{3}{2}\left(1+\omega_{de} \Omega_{de} \right)H^2R_h^2
\end{equation}
In a de Sitter universe in which the apparent horizon and 
event horizons coincide such that $R_h=H^{-1}$, the equation of state parameter $\omega_{de}=-1$ and $\Omega_{de}=1,$ the above condition reduces to $HR_h \geq 1.$

Using the relation for $q$ in terms of equation of state and dark energy mass parameter, the above equation can be translated in to the form
\begin{equation}
 q \leq -1 + {\dot{R_h} \over H^2 R_h^2}
\end{equation}
For de Sitter $\dot{R_h}=0$, so the above condition critically satisfied with $q=-1$ and the GSL is satisfied. In such a universe the equation of state $\omega_{de}=-1.$

For the holographic dark energy described in section 2, it was found that for positive values of $\beta$ the equation of state $\omega_{de} >-1,$ 
and for -ve values of $\beta$ the equation of state, $\omega_{de} \leq -1.$ 
We made a numerical analysis on the evolution of the condition (\ref{eqn:horcond1}) by plotting $HR_h-1- (3/2)(1+\omega_{de} \Omega_{de})H^2 R_h^2$ with the redshift $z.$ The plots shows 
that for positive values of $\beta$ parameter (see figure \ref{fig:ehpob}) the GSL is partially satisfied at the event horizon on the other hand for negative values (see figure \ref{fig:ehneb}) 
the GSL is completely violated at the 
event horizon.

\section{GSL under thermal non-equilibrium condition}

For the further analysis we consider the universe with apparent horizon as the boundary.
A complete thermal non-equilibrium condition means that there is no
interaction between the dark sectors hence they have different
temperatures that is,$T_{de}\neq T_{m} \neq T_h$ 
where $T_{de}$ is the temperature of the dark
energy, $T_{m}$ is the temperature of the dark matter ,and $T_{h}$
is the temperature of the horizon. Using the Gibb's relation the rate of change of the total entropy comprising that of entropy of the dark energy, dark matter and horizon, 
can be written as,
\begin{equation}
 S^{'}= H^{-1} \left[ {4\pi \over H^2} \left({\rho_{de}+P_{de} \over T_{de} } + {\rho_m \over T_m} \right) q + {8\pi \over T_h} (1+q) \right] 
\end{equation}
which can be simplified using the basic definition of $q$ and Friedmann equations to 
\begin{equation} \label{eqn:sfull1}
 S^{'}=H^{-1} \left[ 8\pi (1+q) \left({q \over T_{de}} + \frac{1}{T_h} \right) + 12\pi q \Omega_m \left(\frac{1}{T_m} - \frac{1}{T_{de}} \right) \right].
\end{equation}
As a first case we will consider the validity of the GSL in which the dark sectors were in equilibrium, $T_{de}=T_m.$ Then the second term in the above equation will vanish.
It is clear from the remaining equation that the validity of the GSL demands that $T_{de}/T_h \geq -q$ provided $(q+1) \geq 0.$ This implies that for an accelerating universe 
with $q  \geq-1,$  the dark energy 
temperature is greater than horizon temperature, $T_{de}>T_h.$ For the  special case $q=-1$, corresponds to de Sitter type universe in which
 $H \propto \sqrt{\Lambda}$, where $\Lambda$ is the cosmological constant, the change in entropy is zero as a 
result the total entropy of the universe remains 
constant. For $(1+q)<0$, the GSL is satisfied provided $T_{de}/T_h <-q.$ The condition $(1+q)<0$ implies that the universe is in the phantom phase \cite{Cald1} and for 
such phases of expansion the above equation shows that the dark energy temperature is less than the dark matter temperature.
\begin{figure}[ht]
\centering
\includegraphics[scale=1]{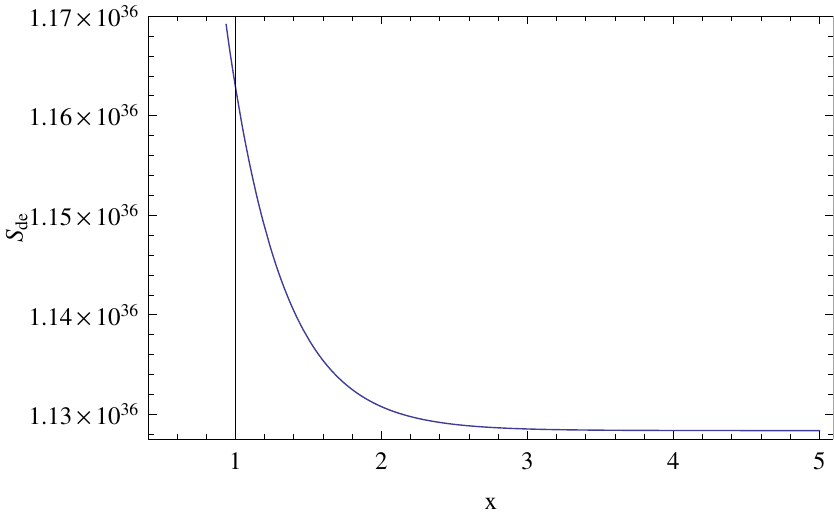}
\caption{variation of entropy of the holographic Ricci dark energy
against $x$ for $k$=1.25 under thermal non-equilibrium condition}
\label{fig:Sde-x.pdf}
\end{figure}

As a further case let us now avoid the dark matter contribution to the total entropy of the universe. The equation (\ref{eqn:sfull1}) now become,
\begin{equation}
 S^{'}=8\pi (1+q) \left({q \over T_{de}} + \frac{1}{T_h} \right) - {12\pi q \Omega_m \over T_{de}}.
\end{equation}
From this relation it can be shown that the GSL is satisfied if 
\begin{equation}
 {T_{de} \over T_h} > -{3(1+\omega_{de})\Omega_{de} \over 2} \left({q \over 1+q} \right)
\end{equation}
\begin{figure}[ht]
\centering
\includegraphics[scale=1]{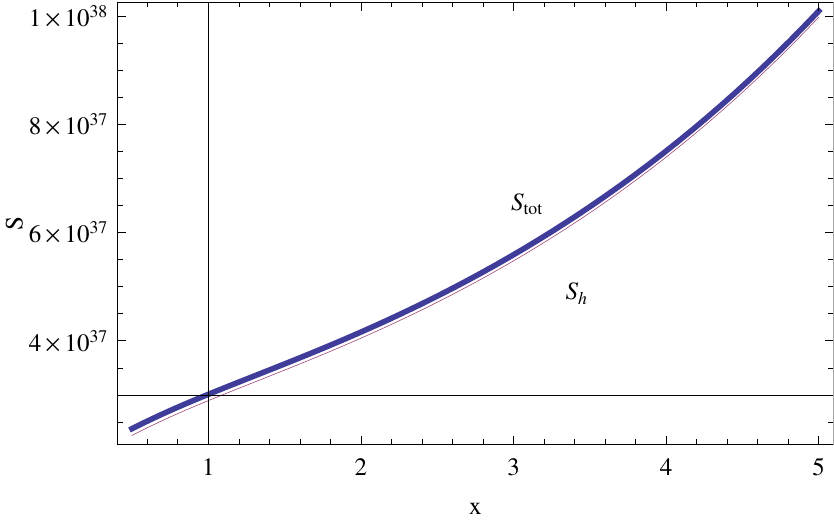}
\caption{ Variation of the sum of entropy of the
holographic Ricci dark energy and the apparent horizon along with
the entropy of the horizon alone against $x$ for $k$=1.25 under
thermal non-equilibrium condition} 
\label{fig:entropy2}
\end{figure}
In an accelerated expanding universe, i.e. $q<0$ and if the expansion is such that the universe will not enter the phantom behavior so that $(1+\omega_{de})>0$ and $(1+q)>0$
then the above ratio implies that $T_{de} / T_h >0,$  that is the dark energy temperature is greater than that of the horizon. In the cases of $(1+\omega_{de})<0$ and $(1+q)<0$
corresponds to the phantom phase also the $T_{de} > T_h.$ So in a dark energy dominated universe the temperature of the dark energy is greater than the temperature of the dark matter 
for the GSL to be valid.

Let us now assume that the temperature of the 
dark energy, $T_{de} \propto T_h,$ especially assume that $T_{de}=k T_h$ with $k>1.$ Then the entropy of the dark energy is
\begin{equation}
 S_{de} = \frac{8\pi^{2}}{k H^{2}}\Omega_{de} \left(1+\omega_{de}\right),
\end{equation}
where we have used the standard relation for entropy $S=(\rho + P)V/T$ \cite{Kolb1}. The evolution of $S_{de}$ with $x$ is shown in figure \ref{fig:Sde-x.pdf} and it 
is clear that the entropy of the dark energy is decreasing as the universe expands. The total entropy of the universe, which can be obtained by adding the horizon entropy 
to the above equation, we get (avoiding the matter contribution),
\begin{equation}
 S = \frac{8\pi^{2}}{k H^{2}}\Omega_{de} \left(1+\omega_{de}\right) + \frac{8\pi^2}{H^2}
\end{equation}
The evolution of this total entropy is plotted in figure \ref{fig:entropy2}, shows that the total entropy is increasing as the universe expands, and thus satisfying the GSL. 
This shows that the decrease in the entropy of the dark energy is compensated by the increase in the horizon entropy. This confirms our previous analyses that 
 if the temperature of the dark energy is greater than the horizon temperature, the total entropy of the universe will always increases which guarantees 
the validity of the GSL.

\section{Conclusion}

In this paper we have analysed the GSL in a flat universe with holographic dark energy and non-relativistic dark matter. The analysis were done both under equilibrium 
 and non-equilibrium conditions. In equilibrium conditions the following are our main conclusions. In the case of taking apparent horizon as the boundary of the universe,
 we have obtained that rate of change of total entropy is proportional to 
$(1+q)^2,$ hence the GSL is infact valid for any kind of dark energy. We have verified it with holographic dark energy with Ricci scalar as the IR cut-off. So it can be concluded 
that the apparent horizon is a perfect thermodynamic boundary. We have also analysed the status of the GSL with event horizon as the boundary and found that in the presence of 
holographic dark energy the GSL is only partially satisfied for positive values of the model parameter $\beta$ but completely unsatisfied for negative values of the parameter.
In literature there are studies regarding the status of the GSL at the apparent horizon. By considering viscous dark energy in non-flat universe, it was found in reference 
\cite{Karami1} that the GSL is valid at the apparent horizon. In reference \cite{Ujjal1}, the authors analyzed the validity of the GSL at the apparent horizon with logarithmic corrected 
entropy relation for the horizon and showed that the GSL is satisfied at the boundary if the model parameter $\alpha=0.$ and also holds for $\alpha<0$ if $\dot{H}<0.$ So the validity 
of the GSL at the apparent horizon seems to be a general fact as advocated by our analysis. there are other studies in the literature regarding the status of the GSL at the 
event horizon. For example in references \cite{Ujjal1,Ujjal2}, the authors argued that the GSL is only partially satisfied at the event horizon of the universe.

We have extended our analysis of the validity of the GSL with thermal non-equilibrium conditions. If there is partial non-equilibrium, such that the temperature of the dark sectors 
are equal and is different from the horizon temperature, the validity of the GSL demands that the temperature of the dark sector is greater than the horizon temperature if the 
expansion is the quintessence phase. On the other hand if the universe is in the phantom phase of expansion the dark sector temperature is less than the horizon temperature. But in a
dark energy dominated universe (avoiding the dark matter contribution), equivalent to be a condition of full non-equilibrium where all the components have different temperatures, the 
temperature of the dark energy is greater than the dark matter temperature both in quintessence and phantom phases of expansion. There are attempts to study the thermodynamics of the 
horizon under non-equilibrium conditions. For example in reference \cite{Ujjal2}, the authors have anlaysed the conditions for the validity of the GSL under non-equilibrium condition, 
however no specific conclusions or criterion were obtained for the validity of the GSL. So it seems difficult to analyses the status of GSL under non-equilibrium conditions. However 
we carried out the analysis by assuming the validity of the GSL at the apparent horizon, and have shown that the GSL is valid if the dark energy temperature is greater than the 
horizon temperature.

\end{document}